\documentclass[a4paper]{article}

\usepackage{INTERSPEECH2022}

\usepackage{amsmath,graphicx}
\usepackage{boldline}
\usepackage{multirow}
\usepackage{subcaption}

\usepackage{caption}
\usepackage{bbm}
\def\wshift{w_\mathrm{shift}}
\def\wsparse{w_\mathrm{sparse}}
\def\wsign{w_\mathrm{sign}}
\def\wbsparse{\mathbf w_\mathrm{sparse}}
\def\Rsparse{\mathcal R_\mathrm{sparse}}

\title{Low-bit Shift Network for End-to-End Spoken Language Understanding}
\name{Anderson R. Avila, Khalil Bibi, Rui Heng Yang, Xinlin Li, Chao Xing, Xiao Chen}
\address{Huawei Noah’s Ark Lab\\
Montreal, QC H3N 1X9, Canada}
\email{$\{$anderson.avila, khalil.bibi, rui.heng.yang1,  xinlin.li1, xingchao.ml, chen.xiao2$\}$@huawei.com}

\begin{document}

\maketitle
\begin{abstract}
Deep neural networks (DNN) have achieved impressive success in multiple domains. Over the years, the accuracy of these models has increased with the proliferation of deeper and more complex architectures. Thus, state-of-the-art solutions are often computationally expensive, which makes them unfit to be deployed on edge computing platforms. In order to mitigate the high computation, memory, and power requirements of inferring convolutional neural networks (CNNs), we propose the use of power-of-two quantization, which quantizes continuous parameters into low-bit power-of-two values. This reduces computational complexity by removing expensive multiplication operations and with the use of low-bit weights. ResNet is adopted as the building block of our solution and the proposed model is evaluated on a spoken language understanding (SLU) task. Experimental results show improved performance for shift neural network architectures, with our low-bit quantization achieving 98.76 \% on the test set which is comparable performance to its full-precision counterpart and state-of-the-art solutions.

\end{abstract}

\section{Introduction}
\label{sec:intro}

With the advent of edge computing \cite{shi2016edge} and Tiny ML \cite{banbury2020benchmarking}, interest in shrinking large deep neural network models has increased. The aim is to reduce computational complexity, allowing their deployment in small devices such as laptops, smartphones, and even less powerful gadgets such as wearables and hearing aids \cite{braun2021towards}. Deep neural networks (DNNs), in fact, have achieved breakthrough results in several areas. The outstanding performance of state-of-the-art models has been possible primarily with the development of larger and more complex architectures \cite{braun2021towards}\cite{courbariaux2015binaryconnect}. Training these models, however, is very demanding in terms of required computational resources which is associated with high energy consumption and carbon emission as discussed in \cite{strubell2019energy}. This leads to undesirable financial and environmental costs that researchers have recently been trying to alleviate.

While such models have been generally deployed in the cloud, motivated by its computing power and data processing efficiency \cite{shi2016edge}, there has been an increasing interest in alleviating the burden of cloud computing. This has several benefits for real-world applications. For instance, as more and more data is being generated at the edge of the network, processing such data on small devices can be more efficient than processing it in the cloud \cite{shi2016edge}. Moreover, aside from mitigating data safety and privacy issues, enabling computing on resource-limited devices allows for latency reduction and minimization of data transmission through the network. This can ultimately benefit Internet of Things (IoT) applications \cite{aslanpour2021serverless}.

\begin{equation}
2^{\tilde{p}x} = \begin{cases} x << \tilde{p}, \text{for } \tilde{p} > 0\\ x >> \tilde{p}, \text{for } \tilde{p} < 0  \text{, }\text{where } \tilde{p} \in \mathrm{Z}\\ x , \tilde{p} = 0 \end{cases}
\label{eq:shift}
\end{equation}

We are particularly concerned with the deployment of end-to-end spoken language understanding (e2e SLU) systems on edge devices. While e2e SLU models can provide high accuracy in high-level hardware, deploying such models on edge and internet of things (IoT) devices is still a challenge \cite{saade2019spoken,rongali2020exploring}. Inspired by recent publications \cite{elhoushi2021deepshift}, we investigate a low-bit multiplication-free neural network. Specifically, we explore the ternary shift network introduced in \cite{li2021s}. The method replaces computationally expensive operations (i.e. multiplication) by bitwise shift operations, which mathematically is the same as multiplying by a power of 2 \cite{you2020shiftaddnet}, as shown in eq. (\ref{eq:shift}). The authors in \cite{li2021s} also introduced a new training strategy that showed to be more effective than other shift network solutions, such as the one presented in \cite{elhoushi2021deepshift}, as it mitigates gradient vanishing and weight sign freezing problems in low-bit shift networks.

%Because of its size, CNNs are difficult to deploy to low-power edge devices or IoT devices. One of the main challenges is that the CNN inference comprises a large amount of floating-point multiplications. The hardware implementation of high precision floating-point multiplications is expensive and power-hungry to compute. 

\begin{figure*}[ht]
\centering
  \includegraphics[width=0.92\linewidth]{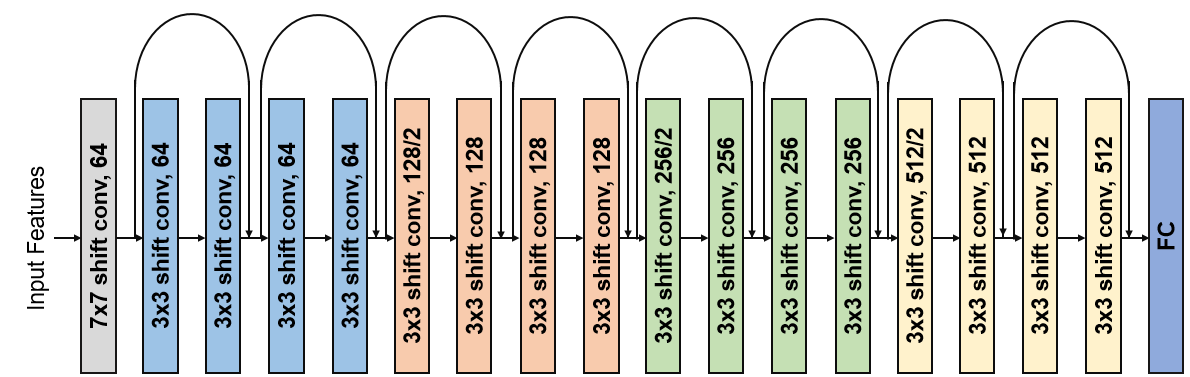}
  \caption{ResNet-18 based on shift operations.}
  \label{fig:s3-3bit}
\end{figure*}

Although quantization is widely used to replace costly high-precision floating-point with low-precision fixed-point representation \cite{bie2019simplified,jacob2018quantization}, it still requires costly multiplication operations. In order to further improve the performance of fixed-point quantization, the proposed ternary shift network decomposes an 3 bits parameter into a sign-sparse-shift 3-fold manner. This way, we attain a low-bit network with weight dynamics that correspond to full-precision networks. As discussed in \cite{you2020shiftaddnet}, bit-shifts can save roughly 196 $\times$  memory and 24 $\times$ energy costs over their multiplication couterpart. 

We use \textit{ResNet} based architecture in our experiments and the models are evaluated on the Fluent Speech Command (FSC) dataset \cite{lugosch2019speech}. Compared to the same architecture with 32 bits, as well as to its quantized version with 16, 8 and 3 bits, experimental results show that our solution can provide accuracy as high as 98 \% for intent classification, thus showing competitive performance to its full-precision counterpart and quantized versions. To the best of our knowledge, this is the first work proposing a multiplication-free low-bit representation as an end-to-end SLU solution.

This paper is organized as follows. Section ~\ref{sec:related} presents the related work. In Section~\ref{sec:s3}, we give details about the Ternary shift network. Section~\ref{sec:slu} introduces the SLU task. In Section~\ref{sec:exp} we present the experimental setup, followed by the result discussion in Section~\ref{sec:results}.

\section{Related Work}
\label{sec:related}

In order to enable the deployment of DNN-based models in resource-constrained scenarios, significant efforts have been made to minimize the computational complexity of neural networks without compromising inference accuracy. Most of these efforts focus on low-precision fixed-point multiplications and to further improve these techniques shift network has been also investigated. In \cite{courbariaux2015binaryconnect}, for instance, the authors introduced a binarization scheme called BinaryConnect. The method replaces multiply-accumulate operations with simple accumulations, allowing to train DNN models with binary weights without compromising the precision of stored weights. In another similar work \cite{rastegari2016xnor}, the authors present an efficient and accurate way of approximating CNNs by binarizing the weights. Two quantized CNNs, namely Binary-Weight-Network and XNOR-Network, are investigated. In the former, CNN filters represented with binary values leads to memory reduction in the order of 32$\times$. Different from the binary approach, where weights are constrained between two values (e.g., 0 and 1), ternary weight networks (TWN) are introduced in \cite{li2016ternary}. The weights in this work are constrained to three values: -1, 0, and 1. According to the authors, their model offers higher accuracy and higher model compression rate (i.e., 16 $\times$, 32$\times$), requiring fewer multiplications compared to its full precision version \cite{li2016ternary}. In a more recent work \cite{elhoushi2021deepshift}, the authors introduced DeepShift. The solution presents two new operations, namely convolutional shift and fully-connected shift, that minimize the number of multiplications by introducing bitwise shift and sign flipping during both training and inference. During inference, for instance, only 5 bits (or less) are required to represent the weights. 

While memory efficient and hardware friendly, training shift networks is challenging. The authors in \cite{li2021s}, pointed out two major problems. First is the performance sensitivity to weight initialization which can lead to accuracy degradation. The second problem is the fact that quantizers are non-differentiable functions and gradient approximators may lead to gradient vanishing and weight sign freezing. To mitigate these problems the ternary shift network, proposed in \cite{li2021s}, is discussed next.

\section{Ternary Shift Neural Networks and S$^3$ Re-parameterization}
\label{sec:s3}

Shift neural network \cite{gudovskiy2017shiftcnn, elhoushi2021deepshift, zhou2017incremental} is a special fixed-point quantized CNN that restricts weight values to the range of zero or positive/negative power of two numbers, thus $\wshift \in \{0\} \cup \{\pm 2^p\}$. The multiplication between a fixed-point integer and a power of two can be implemented with the bit-shift operator. The bit-shift operator's power consumption and hardware area cost are significantly lower than that of the multiplication operator under the same precision. Therefore, Shift networks outperform fixed-point quantized CNN in terms of energy consumption and computational efficiency. Gudovskiy et al. \cite{gudovskiy2017shiftcnn} show that shift networks can achieve 4x energy saving and 2.5x hardware resources saving compared to fixed-point quantized networks on FPGA.

Despite the competitive inference performance of Shift networks, training has been a challenge. All previous training methods \cite{gudovskiy2017shiftcnn, elhoushi2021deepshift, zhou2017incremental} cannot achieve the same accuracy as the full-precision baselines on the ImageNet classification task and require initialization from a pre-trained full-precision checkpoint. Recently, Li et al. \cite{li2021s} proposed S3 re-parameterization, a new method for training low-bit shift networks to tackle this challenge. This method re-parameterizes the discrete weights of shift networks to the product of multiple binary parameters. The low-bit shift networks trained with S3 re-parameterization achieve the same accuracy as regular full-precision CNN in the ImageNet classification task with \textit{ResNet} based architecture. Moreover, unlike previous methods, the S3 method achieves high accuracy without using complex initialization or training strategies.

In this work, the weight value range limits to $\{0, \pm 1, \pm 2, \pm 4\}$. With S3 training method, the low-bit discrete weights are re-parameterized into one sign parameter $\wsign$, one sparse parameter $\wsparse$, and two shift parameters $w_{s1}$, $w_{s2}$. During forward propagation, all parameters are binarized by the Heaviside function $\mathbbm{1}(\cdot)$, and calculate the discrete 3-bit shift network weight values based on eq. (\ref{eq:shift-3bit}). During backward propagation, the straight-through estimator \cite{bengio2013estimating} is utilized to approximate the derivative of the Heaviside function.

\begin{equation} \label{eq:shift-3bit}
\begin{aligned}
    \wshift &= 2^{S} \mathbbm{1}(\wsparse) \{2\mathbbm{1}(\wsign)-1\} \\
    S &= \mathbbm{1}(w_{s2})\{\mathbbm{1}(w_{s1})+1\}
\end{aligned}
\end{equation}

Following the S3 training method, the dense weight regularizer applies to the sparse parameter $\wbsparse$ during training.

\begin{equation} \label{eq:l0_reg}
    \Rsparse(\wbsparse) =  \lVert \max(-\wbsparse,0) \rVert_{1}\\
\end{equation}

%\begin{table}
%\centering
%\caption{Details of the adopted ResNet-18 and ResNet-50.}
%\scalebox{.99}{
%\begin{tabular}{c|c|c}\cline{1-3}
%                          & ResNet - 18              & ResNet - 50            \\ \cline{1-3}
%conv1                     & \multicolumn{2}{c}{7x7,64, stride = 2}            \\ \cline{2-3}
%                          & \multicolumn{2}{c}{3x3 max pool, stride 2}        \\ \cline{1-3}
%conv2\_x                  &  $\begin{bmatrix} 3 \times 3, 64\\ 3 \times 3, 64\\ \end{bmatrix} \times 3$   &   $\begin{bmatrix} 1 \times 1, 64\\ 3 \times 3, 64\\ 1 \times 1, 256 \\ \end{bmatrix} \times 3$                       \\  \cline{1-3}
%conv3\_x                  &  $\begin{bmatrix} 3 \times 3, 128\\ 3 \times 3, 128\\ \end{bmatrix} \times 3$         &   $\begin{bmatrix} 1 \times 1, 128\\ 3 \times 3, 128\\ 1 \times 1, 512 \\ \end{bmatrix} \times 4$                      \\  \cline{1-3}
%conv4\_x                  &  $\begin{bmatrix} 3 \times 3, 256\\ 3 \times 3, 256\\ \end{bmatrix} \times 3$         &   $\begin{bmatrix} 1 \times 1, 256\\ 3 \times 3, 256\\ 1 \times 1, 1024 \\ \end{bmatrix} \times 6$                       \\  \cline{1-3}
%conv5\_x                  &  $\begin{bmatrix} 3 \times 3, 512\\ 3 \times 3, 512\\ \end{bmatrix} \times 3$         &   $\begin{bmatrix} 1 \times 1, 512\\ 3 \times 3, 512\\ 1 \times 1, 2048 \\ \end{bmatrix} \times 3$                       \\  \cline{1-3}
%                          & \multicolumn{2}{c}{average pool, N-d fc, softmax} \\ \cline{1-3}
%\end{tabular}}
%\label{arch:resnet}
%\end{table}

\section{Spoken Language Understanding}
\label{sec:slu}
Voice command recognition enables users to express their intentions via their voices. This is important for modern AI human computer interaction as it allows voice control of smart homes, smart speakers, and other devices such as phones and tablets. Recently, researchers proposed the end-to-end spoken language understanding~\cite{lugosch2019speech} to extract the structure representation of user's intention directly from the speech signals. An end-to-end SLU aims at classifying the observed utterance into one of the predefined semantic classes $L = \{l_1,...,l_k\}$  \cite{masumura2018neural}. Thus, a semantic classifier is trained to maximize the class-posterior probability for a given observation, $X = \{x_1, x_2,..., x_j\}$, representing a sequence of acoustic frames. This is achieved by the following probability:

\begin{equation}
\label{eq:suc_1}
    L^* = \arg\max_{L} P(L|X, \theta)
\end{equation}

\begin{figure}
\centering
  \includegraphics[width=0.99\linewidth]{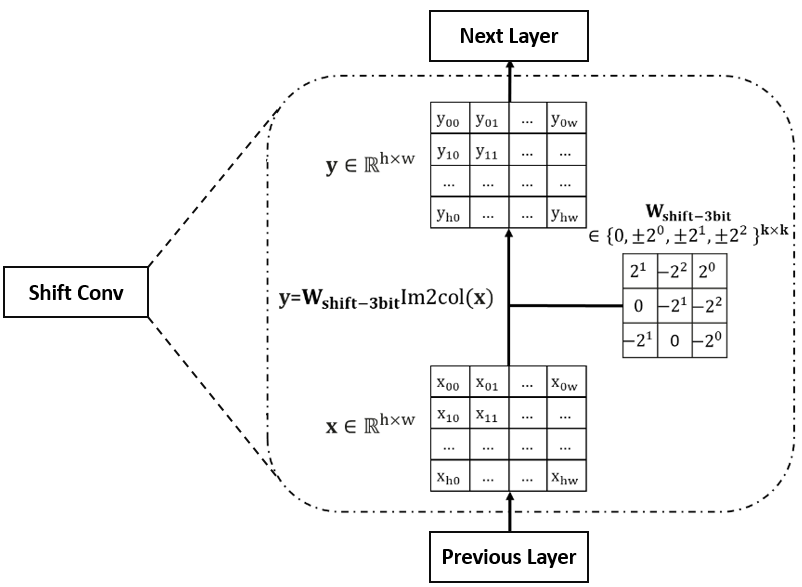}
  \caption{3-bit shift network with S3 re-parameterization.}
  \label{fig:shiftconv}
\end{figure}

\noindent where $\theta$ is the parameters of the end-to-end neural network model. Compared to the tandem ASR + NLU architecture, the end-to-end approach solves the so-called ASR error propagation problem, but usually requires large amounts of data to deal with the high variability present in the speech signal.

\section{Experimental Setup}
\label{sec:exp}

\subsection{Dataset}

The FSC dataset comprises single-channel audio clips sampled at 16 kHz. The data was collected using crowdsourcing, with participants requested to cite random phrases for each intent twice. It contains about 19 hours of speech, providing a total of 30,043 utterances cited by 97 different speakers. The data is split in such a way that the training set contains 14.7 hours of data, totaling 23,132 utterances from 77 speakers. Validation and test sets comprise 1.9 and 2.4 hours of speech, leading to 3,118 utterances from 10 speakers and 3,793 utterances from other 10 speakers, respectively. The dataset comprises a total of 31 unique intent labels resulted in a combination of three slots per audio: action, object, and location. The latter can be either “none”, “kitchen”, “bedroom”, “washroom”, “English”, “Chinese”, “Korean”, or “German”. In our experiments, we defined intent as the combination of action and object, which led to a total of 15 different intent labels. Location was defined as slot, which led to a total of 8 different slot labels. More details about the dataset can be found in \cite{lugosch2019speech}.

\subsection{Features}

In this work, each speech signal is (re)sampled at 16 kHz. We then extract 80-dimensional log Mel-Filterbank features. To extract the Mel features, the audio signal is processed in frames of 320 samples (i.e., 20-ms window length),  with a step  size of 160 samples  (that  is,  10-ms  hop-size). Global Cepstral Mean
and Variance Normalization (CMVN) are applied in order to mitigate the mismatch between training and testing data. After extracting mel-filterbank features, a stacking operation is performed. Specifically, we stacked 5 input frames with stride 2.

\begin{figure}
\centering
  \includegraphics[width=0.99\linewidth]{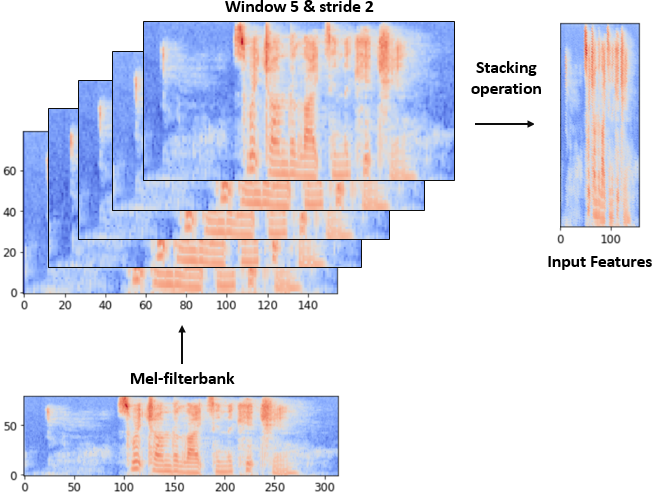}
  \caption{Stacking operation applied after feature extraction.}
  \label{fig:extraction}
\end{figure}

\subsection{Baselines}
The proposed method is compared to several baselines, including 32-bits full-precision, integer quantization at 16, 8, 4, 3 and 2 bits, all based on kernel multiplication. These baselines are referred to as \textit{FP32}, \textit{Q16}, \textit{Q8}, \textit{Q4}, \textit{Q3} and \textit{Q2}. We also included Deepshift as another shift network solution as baseline \cite{elhoushi2021deepshift}, and we refer to this solution as \textit{D4}, \textit{D3}, \textit{D2} for quantization at 4, 3 and 2 bits. The proposed shift network is named \textit{S4}, \textit{S3} and \textit{S2}, also for quantization at 4, 3 and 2 bits. We also included as baseline the work proposed in \cite{lugosch2019speech}, in which the SLU model is first pre-trained to predict phonemes on the first layers and words, thus optimizing the feature representation for SLU.

ResNet-18 is used as our backbone and they are trained from scratch. The architecture adopted in this work is based on the model proposed in \cite{he2016deep}, with the adaptation of having only one input channel as our input representation is based on mel-filterbank.  

%We use ResNet-18 and ResNet-50 as our backbone.

\begin{figure*}[ht]
\centering
\begin{subfigure}{.45\textwidth}
  \centering
  \includegraphics[width=.99\linewidth]{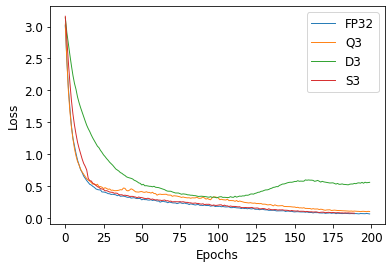}
  \caption{}
  \label{fig:sub1}
\end{subfigure}%
\begin{subfigure}{.5\textwidth}
  \centering
  \includegraphics[width=.91\linewidth]{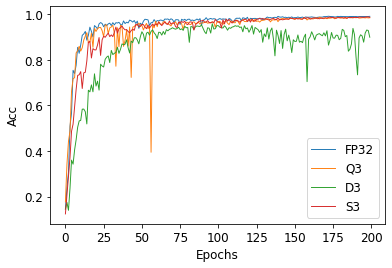}
  \caption{}
  \label{fig:sub2}
\end{subfigure}
\caption{Performance in terms of (a) loss and (b) accuracy of the ResNet-18 explored in different versions, including its 32-bits full-precision (\textit{FP32}), integer quantization (\textit{Q3}), Deepshift (\textit{D3}) and proposed ternary shift network (\textit{S3}), all quantized at 3 bits.}
\label{fig:plots_acc}
\end{figure*}

\subsection{Experimental Settings}

Our network is trained on mini-batches of 8 samples over a total of 200 epochs. In \cite{li2021s}, for instance, the authors pointed out the importance of training the model for 200 epochs. In our experiments, we found Adam optimizer converging faster, but providing poorer generalization performance compared to the SGD optimizer. Thus, except for DeepShift, trained with the Adam optimizer, all models were trained using the SGD optimizer. The initial learning rate was set to $0.0001$ and cosine annealing was also applied. The FSC dataset is split into training, validation, and test sets, and the hyperparameters and final model were selected based on the performance of the validation set. All reported results are based on the accuracy of validation and test sets.

%Our model was trained using either the Adam optimizer \cite{kingma2014adam} or SGD \cite{duda2019sgd}.

\section{Results}
\label{sec:results}

Figure~\ref{fig:plots_acc}-a and Figure~\ref{fig:plots_acc}-b show the performance of 3 baselines (i.e. \textit{FP32, Q3, D3}) and the proposed low-bit shift network, referred to as \textit{S3}, in terms of loss and accuracy, respectively. The losses are attained from the train set whereas the accuracy is based on the test set. We can observe that the quantized models, \textit{Q3, S3}, follow similar trend as the full-precision model regarding the loss and their accuracy with \textit{S3} offering slightly better results. While these models have a monotonic decay of their losses and an increasing accuracy towards the end of training, \textit{D3} enters an overfitting regime around epoch 100.

%In the case of the proposed model (i.e. S3), it is due to the increased number of parameters required by the shift operation during training. All quantized models, however, follow the same trend as its full-precision counterpart. The plots also show that SGD seems to help the model to converge faster compared to the Adam optimizer. In Fig.~\ref{fig:plots_acc}-b, we can observe that the accuracy is not compromised by the low-precision representation which is the main focus of this work. 

In Table~\ref{tab:results18}, we present the performance in terms of accuracy and other quantization settings are explored. Results are based on the best model and not on the final model. For \textit{D3}, for instance, we chose a model around epoch 100 which gave the best accuracy on the validation set. The same was done for the other experiments in Table~\ref{tab:results18}. Results are compatible to the baseline results presented in \cite{lugosch2019speech}, with the full-precision (\textit{FP32}) even outperforming it with accuracy as high as 92.34 \% and 98.97 \% for the validation and test set, respectively, followed by our proposed solution quantized with 4 bits, \textit{S4}, which provides 92.24 \% and 98.76 \%, respectively, for the validation and test sets as well.

Quantization had a mild impact on the overall performance. For integer quantization, for example, quantizing with 16 bits showed to be slightly benefitial compared to quantizatizing with 2 bits with latter dropping accuracy only 0.06 \% on the test set. Similar trend was found for the ternary shift network. While quantizing with 4 bits gives 98.76 \% accuracy, quantizing with 2 bits provides 98.41 \% accuracy, representing a decay of only 0.35 \%. These results show that compressing a model with low-bit representation is feasible for the SLU task and with the ternary shift network fast inference will be possible with the absence of expensive multiplication operations.

%The results achieved by the proposed model are comparable to the full-precision network, but slightly lower. For we also notice that ResNet-18 seems to be the optimal architecture for both full-precision and low-bit networks investigated in this work. We hypothesis that ResNet-50 is over-parameterized for the given task and dataset. In fact, recent research has shown that a simpler models can be as effective as complex and task specific architectures \cite{tolstikhin2021mlp}. Note that our results are compatible with the ones presented by the authors that introduced the FSC dataset. In their work~\cite{lugosch2019speech}, accuracy was reported as high as 98.80 \% on the test set, using the full training data and unfreezing the pre-trained layers. Note that the authors in that work consider 31 intents in their settings.

\begin{table}
	\centering
	\caption{Experimental results for ResNet-18. Performance is reported in terms of accuracy (\%).}
	\scalebox{0.87}{\begin{tabular}{llccc}\
		Kernel Operation & Method & \# Bits & Val & Test\\
		\hline
		- & \cite{lugosch2019speech}      & -  & 89.50 & 98.80\\
        \hline
		\multirow{6}{*}{Multiplication}   & FP32      & 32 & \textbf{92.36} & \textbf{98.97}\\
        & Q16      & 16  & 90.73 & 98.47\\
        & Q8       & 8  & 90.69 & 98.57\\
        & Q4       & 4  & 91.24 & 98.52\\
        & Q3       & 3  & 91.43 & 98.28 \\
        & Q2       & 2  & 91.08 & 98.39 \\
        \hline
         \multirow{6}{*}{Shift}                & D4        & 4  & 79.76 & 93.16 \\
                        & D3 & 3  & 82.39 & 94.83\\                     
                        & D2 & 2  & 80.46 & 93.80 \\                     
                        & S4 (ours) & 4  & 92.24 & 98.76\\                     
                         & S3 (ours) & 3  & 90.31 & 98.41\\                     
                         & S2 (ours) & 2  & 90.66 & 98.41\\
        \hline
		%\clineB{1-7}{1.5}
	\end{tabular}}
    \label{tab:results18}	
\end{table}

%\begin{table}
%	\centering
%	\caption{Experimental results for ResNet-50. Performance is reported in terms of accuracy (\%).}
%	\scalebox{0.87}{\begin{tabular}{llccc}\
%		Kernel Operation & Method & \# Bits & Val & Test \\
%		\hline
%		\multirow{4}{*}{Multiplication}   & FP32      & 32 & 92.01* & 98.83*\\
                         
%        & Q16       & 16  & 91.62 & 98.81 \\
%        & Q8       & 8  & 91.21 & 98.07 \\
%        & Q3       & 3  & 90.41 & 98.25 \\    \hline 
%        \multirow{2}{*}{Shift} & DS & 3  & 80.59* & 93.30* \\
%                         & S3 (ours) & 3  & 86.49 & 96.91 \\
%        \hline
		%\clineB{1-7}{1.5}
%	\end{tabular}}
%    \label{tab:results50}	
%\end{table}

\section{Conclusion}
\label{sec:prior}

In this work, we propose a low-bit deep neural network for end-to-end (e2e) spoken language understanding (SLU). For that, a power-of-two quantization is applied on continuous weights from a full-precision ResNet-18. The low-bit representation is further decomposed in a sign-sparse-shift 3-fold manner. The method reduces computation complexity by removing expensive multiplication operations and using low-bit weights. We evaluate the performance of the proposed model on the Fluent Speech Command dataset. Our solution can provide accuracy as high as 98.76 \%, showing that it can achieve comparable performance to its full-precision counterpart and state of the art models. As future work, we plan to investigate the use of our solution on recurrent neural networks as they might further improved inference latency for the e2e SLU task.

\bibliographystyle{IEEEtran}

\bibliography{mybib}

\begin{thebibliography}{10}
\providecommand{\url}[1]{#1}
\csname url@samestyle\endcsname
\providecommand{\newblock}{\relax}
\providecommand{\bibinfo}[2]{#2}
\providecommand{\BIBentrySTDinterwordspacing}{\spaceskip=0pt\relax}
\providecommand{\BIBentryALTinterwordstretchfactor}{4}
\providecommand{\BIBentryALTinterwordspacing}{\spaceskip=\fontdimen2\font plus
\BIBentryALTinterwordstretchfactor\fontdimen3\font minus
  \fontdimen4\font\relax}
\providecommand{\BIBforeignlanguage}[2]{{%
\expandafter\ifx\csname l@#1\endcsname\relax
\typeout{** WARNING: IEEEtran.bst: No hyphenation pattern has been}%
\typeout{** loaded for the language `#1'. Using the pattern for}%
\typeout{** the default language instead.}%
\else
\language=\csname l@#1\endcsname
\fi
#2}}
\providecommand{\BIBdecl}{\relax}
\BIBdecl

\bibitem{shi2016edge}
W.~Shi, J.~Cao, Q.~Zhang, Y.~Li, and L.~Xu, ``Edge computing: Vision and
  challenges,'' \emph{IEEE internet of things journal}, vol.~3, no.~5, pp.
  637--646, 2016.

\bibitem{banbury2020benchmarking}
C.~R. Banbury, V.~J. Reddi, M.~Lam, W.~Fu, A.~Fazel, J.~Holleman, X.~Huang,
  R.~Hurtado, D.~Kanter, A.~Lokhmotov \emph{et~al.}, ``Benchmarking tinyml
  systems: Challenges and direction,'' \emph{arXiv preprint arXiv:2003.04821},
  2020.

\bibitem{braun2021towards}
S.~Braun, H.~Gamper, C.~K. Reddy, and I.~Tashev, ``Towards efficient models for
  real-time deep noise suppression,'' in \emph{ICASSP 2021-2021 IEEE
  International Conference on Acoustics, Speech and Signal Processing
  (ICASSP)}.\hskip 1em plus 0.5em minus 0.4em\relax IEEE, 2021, pp. 656--660.

\bibitem{courbariaux2015binaryconnect}
M.~Courbariaux, Y.~Bengio, and J.-P. David, ``Binaryconnect: Training deep
  neural networks with binary weights during propagations,'' in \emph{Advances
  in neural information processing systems}, 2015, pp. 3123--3131.

\bibitem{strubell2019energy}
E.~Strubell, A.~Ganesh, and A.~McCallum, ``Energy and policy considerations for
  deep learning in nlp,'' \emph{arXiv preprint arXiv:1906.02243}, 2019.

\bibitem{aslanpour2021serverless}
M.~S. Aslanpour, A.~N. Toosi, C.~Cicconetti, B.~Javadi, P.~Sbarski, D.~Taibi,
  M.~Assuncao, S.~S. Gill, R.~Gaire, and S.~Dustdar, ``Serverless edge
  computing: vision and challenges,'' in \emph{2021 Australasian Computer
  Science Week Multiconference}, 2021, pp. 1--10.

\bibitem{saade2019spoken}
A.~Saade, J.~Dureau, D.~Leroy, F.~Caltagirone, A.~Coucke, A.~Ball, C.~Doumouro,
  T.~Lavril, A.~Caulier, T.~Bluche \emph{et~al.}, ``Spoken language
  understanding on the edge,'' in \emph{2019 Fifth Workshop on Energy Efficient
  Machine Learning and Cognitive Computing-NeurIPS Edition (EMC2-NIPS)}.\hskip
  1em plus 0.5em minus 0.4em\relax IEEE, 2019, pp. 57--61.

\bibitem{rongali2020exploring}
S.~Rongali, B.~Liu, L.~Cai, K.~Arkoudas, C.~Su, and W.~Hamza, ``Exploring
  transfer learning for end-to-end spoken language understanding,'' \emph{arXiv
  preprint arXiv:2012.08549}, 2020.

\bibitem{elhoushi2021deepshift}
M.~Elhoushi, Z.~Chen, F.~Shafiq, Y.~H. Tian, and J.~Y. Li, ``Deepshift: Towards
  multiplication-less neural networks,'' in \emph{Proceedings of the IEEE/CVF
  Conference on Computer Vision and Pattern Recognition}, 2021, pp. 2359--2368.

\bibitem{li2021s}
X.~Li, B.~Liu, Y.~Yu, W.~Liu, C.~Xu, and V.~P. Nia, ``$\mathcal{S}^3$:
  Sign-sparse-shift reparametrization for effective training of low-bit shift
  networks,'' \emph{arXiv preprint arXiv:2107.03453}, 2021.

\bibitem{you2020shiftaddnet}
H.~You, X.~Chen, Y.~Zhang, C.~Li, S.~Li, Z.~Liu, Z.~Wang, and Y.~Lin,
  ``Shiftaddnet: A hardware-inspired deep network,'' \emph{Advances in Neural
  Information Processing Systems}, vol.~33, pp. 2771--2783, 2020.

\bibitem{bie2019simplified}
A.~Bie, B.~Venkitesh, J.~Monteiro, M.~Haidar, M.~Rezagholizadeh \emph{et~al.},
  ``A simplified fully quantized transformer for end-to-end speech
  recognition,'' \emph{arXiv preprint arXiv:1911.03604}, 2019.

\bibitem{jacob2018quantization}
B.~Jacob, S.~Kligys, B.~Chen, M.~Zhu, M.~Tang, A.~Howard, H.~Adam, and
  D.~Kalenichenko, ``Quantization and training of neural networks for efficient
  integer-arithmetic-only inference,'' in \emph{Proceedings of the IEEE
  conference on computer vision and pattern recognition}, 2018, pp. 2704--2713.

\bibitem{lugosch2019speech}
L.~Lugosch, M.~Ravanelli, P.~Ignoto, V.~S. Tomar, and Y.~Bengio, ``Speech model
  pre-training for end-to-end spoken language understanding,'' \emph{arXiv
  preprint arXiv:1904.03670}, 2019.

\bibitem{rastegari2016xnor}
M.~Rastegari, V.~Ordonez, J.~Redmon, and A.~Farhadi, ``Xnor-net: Imagenet
  classification using binary convolutional neural networks,'' in
  \emph{European conference on computer vision}.\hskip 1em plus 0.5em minus
  0.4em\relax Springer, 2016, pp. 525--542.

\bibitem{li2016ternary}
F.~Li, B.~Zhang, and B.~Liu, ``Ternary weight networks,'' \emph{arXiv preprint
  arXiv:1605.04711}, 2016.

\bibitem{gudovskiy2017shiftcnn}
D.~A. Gudovskiy and L.~Rigazio, ``Shiftcnn: Generalized low-precision
  architecture for inference of convolutional neural networks,'' \emph{arXiv
  preprint arXiv:1706.02393}, 2017.

\bibitem{zhou2017incremental}
A.~Zhou, A.~Yao, Y.~Guo, L.~Xu, and Y.~Chen, ``Incremental network
  quantization: Towards lossless cnns with low-precision weights,'' \emph{arXiv
  preprint arXiv:1702.03044}, 2017.

\bibitem{bengio2013estimating}
Y.~Bengio, N.~L{\'e}onard, and A.~Courville, ``Estimating or propagating
  gradients through stochastic neurons for conditional computation,''
  \emph{arXiv preprint arXiv:1308.3432}, 2013.

\bibitem{masumura2018neural}
R.~Masumura, Y.~Ijima, T.~Asami, H.~Masataki, and R.~Higashinaka, ``Neural
  confnet classification: Fully neural network based spoken utterance
  classification using word confusion networks,'' in \emph{2018 IEEE
  International Conference on Acoustics, Speech and Signal Processing
  (ICASSP)}.\hskip 1em plus 0.5em minus 0.4em\relax IEEE, 2018, pp. 6039--6043.

\bibitem{he2016deep}
K.~He, X.~Zhang, S.~Ren, and J.~Sun, ``Deep residual learning for image
  recognition,'' in \emph{Proceedings of the IEEE conference on computer vision
  and pattern recognition}, 2016, pp. 770--778.

\end{thebibliography}

% \begin{thebibliography}{9}
% \bibitem[1]{Davis80-COP}
%   S.\ B.\ Davis and P.\ Mermelstein,
%   ``Comparison of parametric representation for monosyllabic word recognition in continuously spoken sentences,''
%   \textit{IEEE Transactions on Acoustics, Speech and Signal Processing}, vol.~28, no.~4, pp.~357--366, 1980.
% \bibitem[2]{Rabiner89-ATO}
%   L.\ R.\ Rabiner,
%   ``A tutorial on hidden Markov models and selected applications in speech recognition,''
%   \textit{Proceedings of the IEEE}, vol.~77, no.~2, pp.~257-286, 1989.
% \bibitem[3]{Hastie09-TEO}
%   T.\ Hastie, R.\ Tibshirani, and J.\ Friedman,
%   \textit{The Elements of Statistical Learning -- Data Mining, Inference, and Prediction}.
%   New York: Springer, 2009.
% \bibitem[4]{YourName17-XXX}
%   F.\ Lastname1, F.\ Lastname2, and F.\ Lastname3,
%   ``Title of your INTERSPEECH 2022 publication,''
%   in \textit{Interspeech 2022 -- 23\textsuperscript{rd} Annual Conference of the International Speech Communication Association, September 18-22, Incheon, Korea, Proceedings, Proceedings}, 2022, pp.~100--104.
% \end{thebibliography}

\end{document}